\documentstyle[twocolumn,twoside,graphicx,aps,prl,floats]{revtex}
\graphicspath{{../Figures/}{.}}
\begin{document} 

\title{A Phase Front Instability in Periodically Forced Oscillatory Systems} 
\author{Christian Elphick}
\address{Centro de Fisica No Lineal y Sistemas Complejos de Santiago, Casilla
17122, Santiago, Chile}

\author{Aric Hagberg\thanks{\tt Email: aric@lanl.gov}}
\address{Theoretical Division and Center for Nonlinear Studies, MSB284,
Los Alamos National Laboratory, Los Alamos, NM 87545}

\author{Ehud Meron\thanks{\tt Email: ehud@bgumail.bgu.ac.il}}
\address{The Jacob Blaustein Institute for Desert Research and the Physics 
Department,\\ Ben-Gurion University, Sede Boker Campus 84990, Israel}

\date{\today}

\twocolumn[\hsize\textwidth\columnwidth\hsize\csname
@twocolumnfalse\endcsname

\maketitle

\begin{abstract}
Multiplicity of phase states within frequency locked bands in periodically
forced oscillatory systems may give rise to front structures separating 
states with different phases.
A new front instability is found within bands where
$\omega_{forcing}/\omega_{system}=2n$ ($n>1$).
Stationary fronts shifting the oscillation phase by $\pi$ lose 
stability below a critical forcing strength and decompose into $n$ 
traveling fronts each shifting the phase by $\pi/n$. The 
instability designates a transition from stationary two-phase patterns to 
traveling $n$-phase patterns.


\end{abstract}


\vspace{-0.4cm}
\vskip2pc]

Periodic forcing of a single oscillator can lead to rich dynamics
including quasiperiodic oscillations, frequency-locked bands ordered 
through the Farey construction, and low-dimensional 
chaos~\cite{FKS:82,ROSS:84,JBB:84,SHL:85}. 
A typical feature of
a periodically forced oscillatory system is the multiplicity of 
phase states within a given frequency-locked band~\cite{Walg:97}. 
This feature becomes
particularly significant in spatially extended systems where phase fronts
separating different phase states may appear. The 
simplest situation arises in a system 
that is forced at twice the natural oscillation
frequency (hereafter the 2:1 band). A phase front (kink)
connecting two uniform states
whose phases of oscillations differ by $\pi$ then exists (hereafter a 
``$\pi$-front''). The stability
properties of this type of front are well 
studied~\cite{CLHL:90,CoEm:1,MaNe:94,EHMM:97}. 
As the strength of forcing is decreased a stationary front 
loses stability to a pair of
counter-propagating fronts through a pitchfork bifurcation. 
The instability, known also as the Nonequilibrium 
Ising-Bloch bifurcation~\cite{CLHL:90,HaMe:94a}, 
is responsible for the destabilization of standing 
waves and the onset of traveling wave phenomena including spiral waves.

The low resonance bands, 2:1 and 3:1, have been studied both 
theoretically~\cite{Walg:97,CLHL:90,CoEm:1,MaNe:94,CoEm:92,KRT:92} and 
experimentally~\cite{POS:97}.
All phase-front solutions in these bands
shift the oscillation phase by the same angle (in absolute value): 
$\pi$ in the 2:1 band and 
$2\pi/3$ in the 3:1 band. 
At higher resonance bands phase fronts 
that shift 
the phase by different angles may coexist, 
for example, $\pi$-fronts and $\pi/2$-fronts in the 4:1 band.
In this Letter we report on a new front instability: Upon decreasing
the forcing strength a $\pi$-front within the 
$\mbox{2}n\mbox{:1}$ band ($n>1$) 
loses stability
and decomposes into $n$ interacting $\pi/n$-fronts. We analyze in detail 
the 4:1 resonance case and bring numerical evidence 
for the existence of this type of instability in higher resonances.

\begin{figure}
\centering\includegraphics[width=2.8 in]{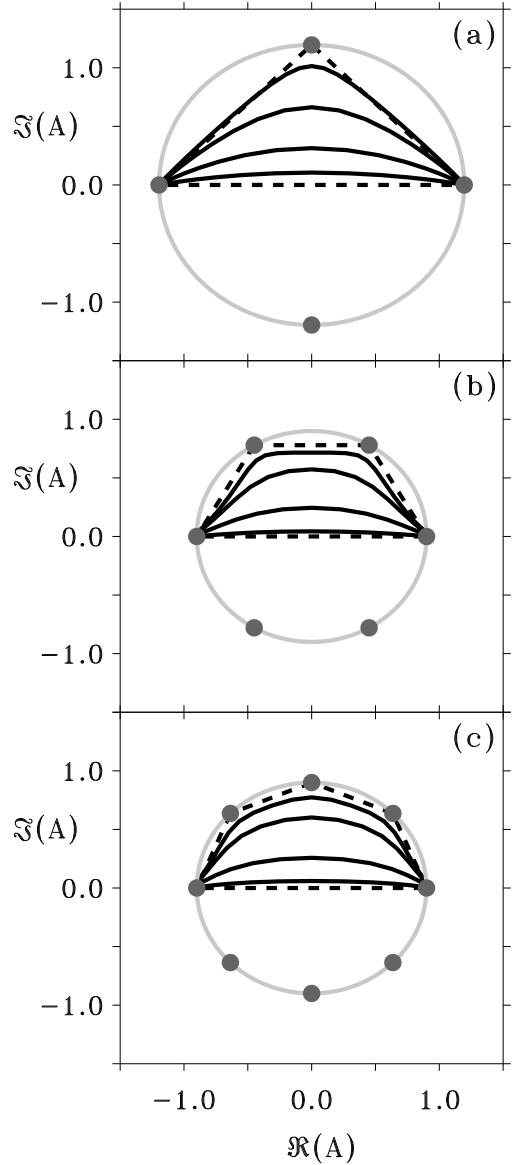}
\caption{Decomposition of $\pi$-fronts into $\pi/n$-fronts within 
$\mbox{2}n\mbox{:1}$  bands for (a) $n=2$, (b) $n=3$, and (c) $n=4$. The dots are the 
uniform phase states along the circle of constant amplitude $|A|$.
The $\pi$-fronts are the dashed lines connecting the states $A_{+1}$ and
 $A_{-1}$ on the $\Re(A)$ axes. The $\pi/n$-fronts are the dashed lines 
connecting the adjacent points along the circle.  The time
evolution of the front decomposition obtained by numerical integration of 
Eqn.~(\protect\ref{hocgl}) is 
shown as the series of solid lines. Parameters: all 
undetermined  
coefficients in Eqn.~(\protect\ref{hocgl}) 
were set to zero except as indicated below.
(a) $\gamma_4=0.3 $, $\mu_4= -1.0$, 
(b) $\gamma_6=0.9$, $\mu_4=-1.0$, $\mu_6=-1.0$,
(c) $\gamma_8=0.75$, $\mu_4=-0.5$, $\mu_6=-0.5$, $\mu_8=-1.0$.
}
\label{fig:pipi2}
\end{figure}

We consider an extended system that is close to a Hopf
bifurcation and externally 
forced with a frequency about four times larger than the Hopf 
frequency. The amplitude of oscillations satisfies the parametrically forced 
complex Ginzburg-Landau (CGL) equation~\cite{Gamb:85,EIT:87}
\begin{eqnarray}
\label{cgl} 
A_t&=& (\mu+i\nu)A+(1+i\alpha)A_{xx}-(1-i\beta)|A|^2A \\
   & & \mbox{} +\gamma_4 {A^{*}}^3\,\nonumber,
\end{eqnarray}
where the subscripts $t$ and $x$ denote partial derivatives with respect to
time and space, $A(x,t)$ is a complex field and 
$\nu,\alpha,\beta$ and $\gamma_4$  are real parameters. We first consider the
gradient version of Eqn.~(\ref{cgl}) by setting
$\nu=\alpha=\beta=0$ and then rescale
time, space and the amplitude $A$ by $\mu$, $\sqrt{\mu/2}$ and $1/\sqrt{\mu}$,
respectively. Keeping the same notations for the scaled quantities the gradient
version takes the form
\begin{equation}
\label{gl} 
A_t= A+\frac{1}{2}A_{xx}-|A|^2A +\gamma_4 {A^{*}}^3\,.
\end{equation}
For $0<\gamma_4<1$ Eqn.~(\ref{gl}) has four stable phase states:
$A_{\pm 1}=\pm\lambda$ and 
$A_{\pm i}=\pm i\lambda$, where $\lambda=1/\sqrt{{1-\gamma_4}}$.
Front solutions connecting pairs of these states divide into two
groups:  
$\pi$-fronts connecting states with a phase shift of $\pi$ 
\begin{eqnarray}
A_{-1\to +1}=A_{+1}\tanh{x}\,,\label{pifronta}\\
A_{-i\to +i}=A_{+i}\tanh{x}\,,\label{pifrontb}
\end{eqnarray}
and $\pi/2$-fronts connecting states with a phase shift of $\pi/2$
(see Fig.~\ref{fig:pipi2}a).
When $\gamma_4=1/3$ the  $\pi/2$-fronts are given by
\begin{eqnarray}
\label{Cfronts}
A_{+1\to +i}=\frac{1}{2}\sqrt{\frac{3}{2}}\bigl[1+i-(1-i)\tanh{x}\bigr]\,,
\nonumber \\
A_{-i\to +1}=\frac{1}{2}\sqrt{\frac{3}{2}}\bigl[1-i+(1+i)\tanh{x}\bigr]\,,
\end{eqnarray}
$A_{-1\to -i}=-A_{+1\to +i}$ and $A_{+i\to -1}=-A_{-i\to +1}$.
Additional front solutions follow from the invariance of Eqn.~(\ref{gl}) under
reflection, $x\to -x$. For example, the symmetric counterparts of 
$A_{+i\to +1}(x)$ and $A_{+1\to -i}(x)$ are $A_{+1\to +i}(x)=A_{+i\to +1}(-x)$
and $A_{-i\to +1}(x)=A_{+1\to -i}(-x)$.

Stability analysis of the $\pi$-fronts (\ref{pifronta}) and (\ref{pifrontb}) shows
that they are stable for $\gamma_4>1/3$. To study the instability at 
$\gamma_4=1/3$ we rewrite Eqn.~(\ref{gl}) in terms of $U=\Re(A)+\Im(A)$ and
$V=\Re(A)-\Im(A)$ :
\begin{eqnarray}
\label{UVeqs}
U_t=U+\frac{1}{2}U_{xx}-\frac{2}{3}U^3 -\frac{d}{2}(U^2-3V^2)U\,,
\nonumber\\
V_t=V+\frac{1}{2}V_{xx}-\frac{2}{3}V^3 -\frac{d}{2}(V^2-3U^2)V\,,
\end{eqnarray}
where 
\[
d=\gamma_4 -1/3\,.
\nonumber
\]
At $d=0$ the two equations decouple  and assume the
solutions $U=\sigma_1 A_0(x-x_1)$ and
$V=\sigma_2 A_0(x-x_2)$, where
$A_0=\sqrt{\frac{3}{2}}\tanh{x}$, $\sigma_{1,2}=\pm 1$ and
$x_1$ and $x_2$ are arbitrary constants. Consider now $d\ne 0$ but small.
The coupling between $U$ and $V$
makes $x_1$ and $x_2$ slow dynamical variables and $U$ and $V$ can be written as
\begin{eqnarray}
U=\sigma_1A_0[x-x_1(t)]+u\,,\nonumber\\
V=\sigma_2A_0[x-x_2(t)]+v\,,
\label{UVsol}
\end{eqnarray}
where $u$ and $v$ are corrections of order $d$. Inserting these 
forms in Eqns.~(\ref{UVeqs}) we obtain
\begin{eqnarray}
{\cal H}_1u &=& \sigma_1\dot x_1A_0^\prime(x-x_1)\nonumber\\ 
&-&\frac{1}{2}d\sigma_1\bigl[A_0^2(x-x_1)-3A_0^2(x-x_2)
\bigr]A_0(x-x_1)\,,\nonumber
\end{eqnarray}
where ${\cal H}_1=-
1-\frac{1}{2}\frac{\partial^2}{\partial x^2}
+2A_0^2(x-x_1)$. A similar equation is obtained for $v$ with the indices 1 and
2 interchanged. Solvability conditions lead to the equation
\begin{eqnarray}
\dot x_1 &=& 
-\frac{27}{16}d \int_{-\infty}^\infty dx \nonumber\\
&&\mbox{}\times\tanh(x-x_1)~{\rm sech}^2(x-x_1)~\tanh^2(x-x_2)
\,,\label{xi}
\end{eqnarray}
and to a similar equation for $x_2$ with the indices 1 and 2 interchanged.
Defining a translational degree of freedom $\zeta=\frac{1}{2}(x_1+x_2)$ and an
order parameter $\chi=\frac{1}{2}(x_2-x_1)$, we obtain from 
(\ref{xi})
\begin{equation}
\dot\zeta = 0\,,\qquad\qquad
\dot\chi = -\frac{27}{16}d J(\chi)\,,\label{chiI}
\end{equation}
where 
\begin{eqnarray}
J(\chi)&=&I(a)= 6(a^{-1}-a^{-3})+(1-3a^{-2})G(a)\,,\nonumber\\
G(a)&=&(1-a^{-2})\ln\Bigr(\frac{1+a}{1-a}\Bigl)
\,,\nonumber
\end{eqnarray}
and $a=\tanh{2\chi}$. Note that Eqns.~(\ref{chiI}) are valid to {\em all} 
orders in $\chi$ and to linear order around $\gamma_4=1/3$. 

\begin{figure}
\centering\includegraphics[width=3.0in]{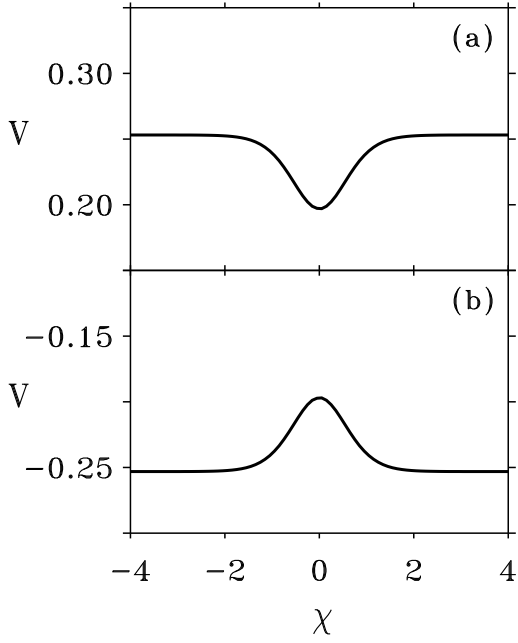}
\caption{The potential $V(\chi)$. (a) For $d>0$
the extremum at $\chi=0$ is a  minimum and and $\chi$ converges to $0$.
(b) For $d<0$ the extremum is a maximum and $\chi$ diverges to $\pm \infty$.
}
\label{fig:potential}
\end{figure}

Figure~\ref{fig:potential}
shows the potential $V(\chi)=-\frac{27}{16}d\int^\chi I(z)dz$ 
associated with Eqn.~(\ref{chiI})
for $d>0$ ($\gamma_4>1/3$) and $d<0$. There is only one 
extremum point, $\chi=0$, of $V$. For $d>0$ it is a minimum and $\chi$
converges to zero. For $d<0$ it is a maximum and $\chi$ diverges to 
$\pm\infty$. To see how the $\chi$ dynamics affect the front solutions
we rewrite the solution form (\ref{UVsol}) in terms of 
the amplitude $A$ and bring it to the form
\begin{equation}
A(x,t)=A_{-i\to +1}(x-x_1)+A_{+1\to +i}(x-x_2) - \lambda +R \,,
\label{interaction}
\end{equation}
where $A_{-i\to +1}$ and $A_{+1\to +i}$ are given in Eqns.~(\ref{Cfronts}),
$R$ is a correction term of order $d$ (related to $u$ and $v$ 
in Eqns.~(\ref{UVsol})) and for concreteness we chose 
$\sigma_1=-\sigma_2=1$. Equation~(\ref{interaction}) describes
two interacting $\pi/2$-fronts centered at $x_1$ and $x_2$. When $d>0$
$\vert\chi\vert$ decreases in time and the two $\pi/2$-fronts attract.
As $\chi\to 0$ (or $x_2\to x_1$) the two $\pi/2$-fronts collapse into 
a single $\pi$-front, $A(x,t)=A_{-i\to +i}(x-x_1) +R$,  given by 
Eqn.~(\ref{pifrontb}). When $d<0$ $\vert\chi\vert$ increases in time
and the two $\pi/2$-fronts repel. Perturbing the unstable $\chi=0$ solution,
the single $\pi$-front decomposes into a pair of $\pi/2$-fronts.
Fig.~\ref{fig:pipi2}$a$
shows phase portraits of the $\pi$ and $\pi/2$-fronts
(dashed lines) and the time evolution of an unstable $\pi$-front for
$d<0$ (solid lines) obtained by numerical integration of Eqn.~(\ref{gl}). 
The approach of the phase portrait to the fixed point $A_{+i}$ on the 
$\Im(A)$ axis 
describes the decomposition into a pair of $\pi/2$-fronts. This behavior 
persists {\em arbitrarily close} to $d=0$ and is related to the absence
of minima in the potential $V(\chi)$ for $d<0$ 
(see Fig.~\ref{fig:potential}$b$). 
At $d=0$ there exists a continuous
family of stationary solutions describing frozen (non-interacting) pairs 
of $\pi/2$-fronts with arbitrary separations $x_2-x_1$. This solution family 
spans the whole phase space inside the dashed triangle 
in Figure~\ref{fig:pipi2}$a$.
Because of the parity breaking symmetry $\chi\to -\chi$ each $\pi$-front
may decompose into one of two pairs of $\pi/2$-fronts with phase portraits
approaching the fixed points $A_{+i}$ and $A_{-i}$.

We extended the derivation of Eqns.~(\ref{chiI}) to the 
nongradient equation (\ref{cgl}), treating the constants
$\nu$, $\alpha$, $\beta$ as small parameters. The $\chi$ equation 
remains unchanged. The $\zeta$ equation takes the form
\begin{equation}
\frac{\sigma_2}{\sigma_1}\dot\zeta=
\nu F_\nu(\chi)+\alpha F_\alpha(\chi)+\beta F_\beta(\chi)\,,
\label{zeta}
\end{equation} 
where 
\begin{eqnarray}
F_\nu&=&-\frac{3}{4}G(a)-\frac{3}{2}a^{-1}\,,\nonumber\\
F_\alpha&=&\frac{3}{4}I(a)\,,\nonumber\\
F_\beta&=&3a^{-1}\bigl(1-\frac{3}{2}a^{-2}\bigr)-\frac{9}{4}a^{-2}G(a)\,.
\nonumber
\end{eqnarray}
Notice that $F_\nu$, $F_\alpha$, and $F_\beta$ are odd functions of $\chi$ and
do not vanish when $d=0$. When $\vert\chi\vert\to\infty$ the right hand side of
(\ref{zeta}) converges to $\frac{3}{2}(\nu+\beta)$, the speed of a 
$\pi/2$-front solution of Eqn.~(\ref{cgl}). The $\chi=0$ solution
(representing a $\pi$-front) remains stationary ($\dot\zeta=0$) in the 
nongradient case as well. At $\gamma_4=1/3$ ($d=0$) it loses stability and
decomposes into a pair of $\pi/2$-fronts which approach the asymptotic speed
$\frac{3}{2}(\nu+\beta)$. Depending on the initial sign of $\chi$ the pair may 
propagate to the left or to the right. 

This behavior is different from that near the
Nonequilibrium Ising-Bloch front bifurcation within the 2:1 band.  In that
case, a stationary Ising front loses stability to a pair of 
counter-propagating Bloch fronts in a pitchfork bifurcation. 
Associated with the bifurcation is a transition from a single-well
potential (Ising front) to a double-well potential (pair of Bloch   
fronts). A comparison with the potentials
in Fig.~\ref{fig:potential} shows the essential difference between the two 
front instabilities. In the 2:1 band the Bloch fronts
approach the Ising front and coincide with it as the distance to the
bifurcation point diminishes to zero. In the 4:1 band, on the other  
hand, the asymptotic solutions  
just below $\gamma_4=1/3$ (the $\pi/2$-front pairs 
as $\vert\chi\vert\to\infty$)
are not smooth continuations of the stationary $\pi$-front (the $\chi=0$
solution) at $\gamma_4=1/3$. 
In particular their speed remains finite ($\frac{3}{2}(\nu+\beta)$) as 
$\gamma_4$ approaches 1/3 from below. At $\gamma_4=1/3$, a whole family of 
propagating solutions appears with speeds ranging continuously from 
$\frac{3}{2}(\nu+\beta)$ to zero 
(pertaining to $\pi/2$-front pairs separated by distances ranging
from infinity to zero).

\begin{figure}
\centering\includegraphics[width=3.15in]{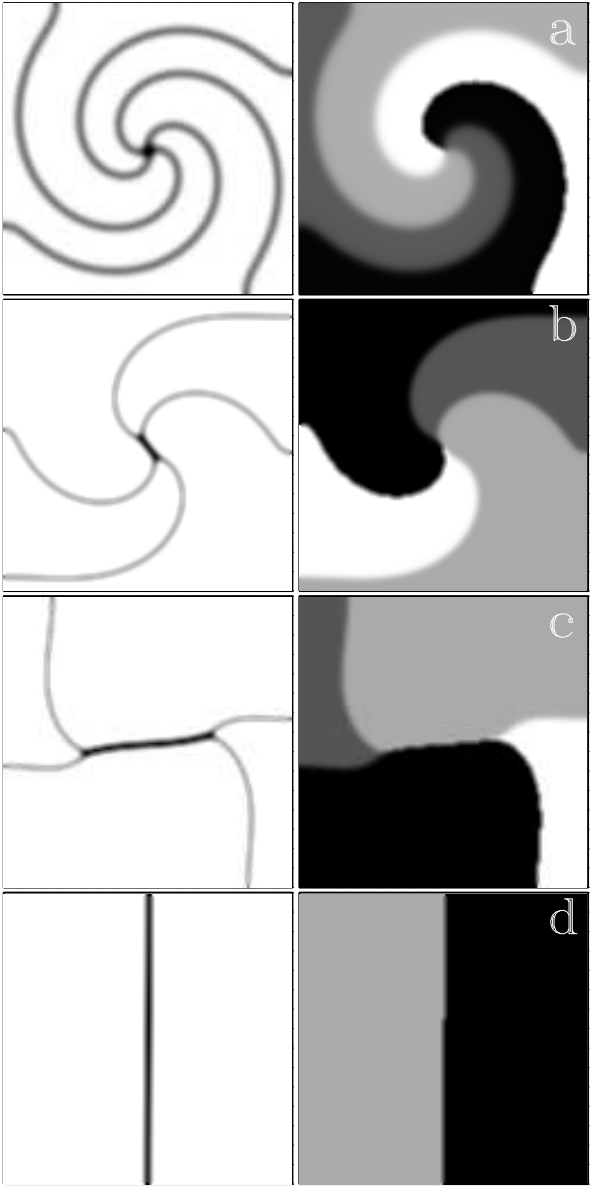}
\caption{Collapse of a rotating four-phase spiral-wave into
a stationary two-phase pattern.  The left column is $|A|$ and
the right column $\arg(A)$.
(a) The initial four-phase spiral wave (computed with $\gamma_4<1/3$).
(b) The spiral core, a 4-point vertex, splits into two 3-point vertices 
connected by a $\pi$-front. 
(c) A two-phase pattern develops as the 3-point vertices further separate. 
(d) The final stationary two-phase pattern.
Parameters: $\gamma_4=0.6,\nu=0.1,x=[0,64],y=[0,64].$
}

\label{fig:spiral}
\end{figure}

The instability of $\pi$-fronts at $\gamma_4=1/3$ ($d=0$) determines
the structure
of stable periodic patterns below and above the instability. In the  
range $\gamma_4>1/3$ {\em two-phase} patterns, involving domains separated by 
$\pi$-fronts, prevail. Below $\gamma_4=1/3$ {\em four-phase} patterns 
dominate. 
Four-phase patterns are not stable for $\gamma_4>1/3$ despite the stability of 
the $\pi/2$-fronts because of the attractive interactions among these
fronts. In the gradient case [Eqn.~(\ref{gl})] all solutions are 
stationary. In the nongradient case [Eqn.~(\ref{cgl})] the two-phase
patterns are stationary while the four-phase patterns propagate. 
Fig.~\ref{fig:spiral}$a$ shows a grey-scale map of a rotating four-phase 
spiral wave for $\gamma_4<1/3$.
Figs.~\ref{fig:spiral}$b,c,d$ show the 
collapse of this spiral wave into a stationary two-phase pattern as 
$\gamma_4$ is increased past $1/3$.
The collapse begins at the spiral core 
where the $\pi/2$-front interactions are the strongest. As pairs of 
$\pi/2$-fronts attract and collapse into $\pi$-fronts, the core splits into
two vertices that propagate away from each other leaving behind 
a two-phase pattern. 


To test whether the instability of $\pi$-fronts exists at higher resonances
we integrated numerically the higher order equation \cite{comment} 
\begin{eqnarray}
\label{hocgl} 
A_t&=& \frac{1}{2}A_{xx}+A+\mu_4|A|^2A+\mu_6|A|^4A +\mu_8|A|^6A\\
   &+& \mbox{} \gamma_4 {A^{*}}^3+\gamma_6 {A^{*}}^5+\gamma_8
   {A^{*}}^7 \nonumber
\,.
\end{eqnarray}
Indeed $\pi$-fronts within the 6:1 and 8:1 bands become unstable as 
$\gamma_6$ and $\gamma_8$, respectively, 
are decreased below a critical value. 
Fig.~\ref{fig:pipi2}$b$ shows the decomposition of a $\pi$-front into three 
$\pi/3$-fronts within the 6:1 band, and Fig.~\ref{fig:pipi2}$c$ shows the 
decomposition into four $\pi/4$-fronts within the 8:1 band. Our conjecture
is that the instability is general, occurring within any
$\mbox{2}n\mbox{:1}$ band ($n>1$). 

The phase front instability and the associated transition from stationary
two-phase patterns to traveling four-phase patterns within the 4:1 band may 
be tested in 
experiments on the ruthenium-catalyzed Belousov-Zhabotinsky reaction
subjected to periodic (in time) uniform illuminations~\cite{POS:97}.

We thank V. Petrov, H.~L. Swinney, M. Clerc, R. Rojas and E. Tirapegui 
for helpful discussions. This research
was supported in part by grant No.~95-00112 from the US-Israel Binational 
Science Foundation (BSF). C.~E. acknowledges the support of the Catedra
Presidencial en Ciencias.


\begin{thebibliography}{10}
\bibitem{comment} No attempt has been made  to include in
Eqn.~(\ref{hocgl}) 
all resonant terms. We have included only terms that appeared
necessary for demonstrating the phase front instability in the higher order 
6:1 and 8:1 bands.
\end{thebibliography}
\end{document}